\documentclass{iaus}
\usepackage{graphicx}
\usepackage{graphicx,bm,url}
\usepackage{natbib}
\graphicspath{{fig/}}

\newcommand{\EQ}{\begin{equation}}
\newcommand{\EN}{\end{equation}}
\newcommand{\EQA}{\begin{eqnarray}}
\newcommand{\ENA}{\end{eqnarray}}

\newcommand{\Fig}[1]{Figure~\ref{#1}}

{}
{}
{}

{}
{}
{}
{}
{}
{}
{}
{}
{}
{}
{}
{}
{}
{}
{}
{}
{}
{}
{}
{}

{}
{}
{}

{}

{}
{}

\newcommand{\xx}{\bm{x}}

\newcommand{\uu}{\mbox{\boldmath $u$} {}}

\newcommand{\nab}{\mbox{\boldmath $\nabla$} {}}
\newcommand{\OO}{\bm{\Omega}}

\def\ga{\mathrel{\mathchoice {\vcenter{\offinterlineskip\halign{\hfil
$\displaystyle##$\hfil\cr>\cr\sim\cr}}}
{\vcenter{\offinterlineskip\halign{\hfil$\textstyle##$\hfil\cr>\cr\sim\cr}}}
{\vcenter{\offinterlineskip\halign{\hfil$\scriptstyle##$\hfil\cr>\cr\sim\cr}}}
{\vcenter{\offinterlineskip\halign{\hfil$\scriptscriptstyle##$\hfil\cr>\cr\sim\cr}}}}}
\def\Ma{\mbox{\rm Ma}}
\def\Co{\mbox{\rm Co}}

\def\St{\mbox{\rm St}}

\def\Rey{\mbox{\rm Re}}

\def\Co{\mbox{\rm Co}}

\def\csz{c_{\rm s0}}
\def\cs{c_{\rm s}}

\def\kf{k_{\rm f}}

\def\kom{k_{\omega}}

\def\urms{u_{\rm rms}}

\newcommand{\yaraa}[3]{ #1, {ARA\&A,} {#2}, #3}

\newcommand{\ymn}[3]{ #1, {MNRAS,} {#2}, #3}

\newcommand{\pproc}[4]{ #1, in {#2}, ed.\ #3 (#4), (in press)}

\title[How can vorticity be produced in irrotationally forced flows?] 
{How can vorticity be produced in irrotationally forced flows?}

\author[Fabio Del Sordo \& Axel Brandenburg]  
{ Fabio Del Sordo
 and Axel Brandenburg}

\affiliation{NORDITA,
Roslagstullsbacken 23, SE-10691 Stockholm, Sweden; and \\
Department of Astronomy, Stockholm University,
SE 10691 Stockholm, Sweden}

\pubyear{2010}
\volume{274}  
\pagerange{119--126}
\jname{Advances in Plasma Astrophysics}
\editors{A.C. Editor, B.D. Editor \& C.E. Editor, eds.}

\begin{document}

\maketitle

\begin{abstract}
A spherical hydrodynamical expansion flow can be described as the gradient of a potential.
In that case no vorticity should be produced, but several additional mechanisms can drive its production.
Here we analyze the effects of baroclinicity, rotation and shear in the case of a viscous fluid.
Those flows resemble what happens in the interstellar medium. 
In fact in this astrophysical environment supernovae explosion are the dominant flows and, in a first approximation,
they can be seen as spherical.
One of the main difference is that in our numerical study we examine only 
weakly supersonic flows, while supernovae explosions are strongly supersonic.

\end{abstract}

Turbulence in the interstellar medium (ISM) is mainly driven by
supernovae explosions, which are among the most dramatic
events in terms of release of energy.
Those explosions are also very important because they can affect
scales up to $\sim 100$ pc.
Moreover they are able to inject in the ISM enough energy to sustain
turbulent flows with velocities of $\sim 10$ km/s.
It is well known that turbulence is one of the key ingredients to be
taken in account when discussing many astrophysical process -- especially
in the production of magnetic fields.
This is indeed one of our ultimate goals,
even though here we do not take any magnetic field into account.
As a first approximation, a supernova explosion can be regarded
as a purely spherical expansion wave.
Thus, we choose a setup consisting of purely potential forcing:
we simulate spherical expansions, as already done by \cite{MB06}.
For our numerical experiments  we use the \textsc{Pencil Code},
http://pencil-code.googlecode.com/.
We have recently extended this work to include rotation, shear,
and baroclinicity; see \cite{DSB10}.
Here we report on some highlights of their work.

We analyze flows that are only weakly supersonic and use a constant and 
uniform viscosity in an unstratified medium.
In our model we solve the Navier-Stokes equations in the viscous case. 
We consider uniform viscosity in an unstratified medium.
We force our system to be only weakly supersonic and we use
a potential forcing $\nabla\phi$ where $\phi$ is given by
randomly placed Gaussian of radius $R$ around the position
$\xx_{\rm f}(t)$.
We use two different forms for the time dependence of the forcing
position $\xx_{\rm f}$.
In the first case we consider a $\delta$-correlated forcing, that is every
timestep has a $\xx_{\rm f}$ completely independent from the previous.
Then we also study the situation in which the forcing remains constant
during a time interval ${\delta{}t}_{\rm force}$.

Next, we add to the system one of three effects that we want to analyze,
taking into account each of them separately.
We start by considering the action of rotation under isothermal condition.
Under the influence of rotation the system is subject to the action of the
Coriolis force.
That is, we add the term $2\OO\times\uu$ in the evolution equation of velocity.
In our simulations we investigate flows with Reynolds numbers (based on
the wavenumber of the energy-carrying eddies) of up to 150.
The aim of this investigation is to quantify the production of vorticity.
We find that vorticity is indeed produced with both kinds of forcing we
have used for driving the spherical expansion.
Nevertheless the case of $\delta$-correlated forcing seems to be more
prone to spurious production of vorticity that we believe is due to
numerical artifact.

\begin{figure}[t!]\begin{center}
\includegraphics[width=\textwidth]{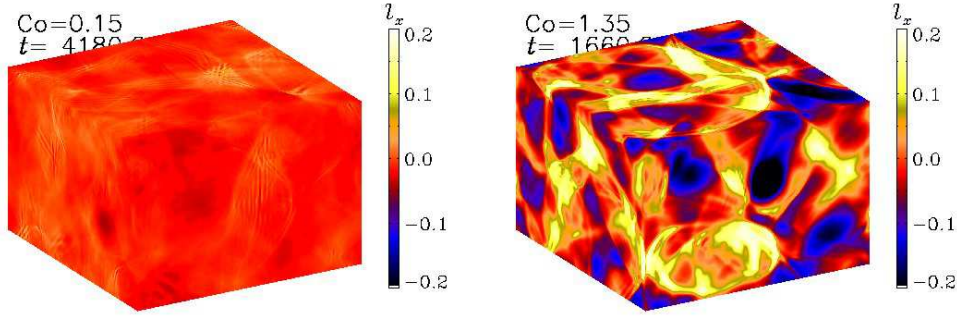}
\end{center}\caption[]{
Vertical component of vorticity on the periphery of the
periodic domain for two values of the Coriolis number.
Note that significant amounts of vorticity are only
being produced when $\Co$ is of the order of unity.
}\label{oo}\end{figure}

\begin{figure}[t!]\begin{center}
\includegraphics[width=.46\textwidth]{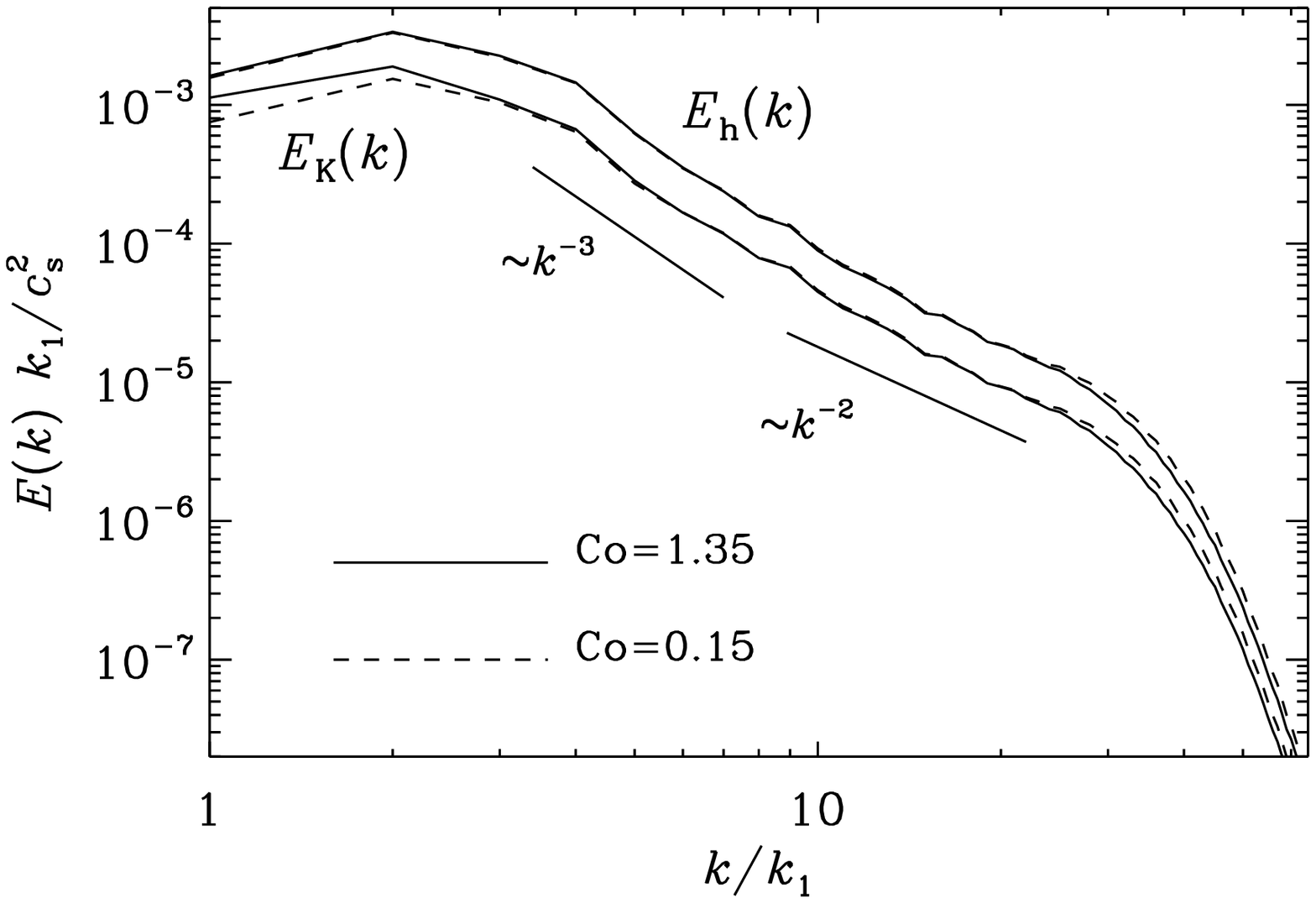}
\includegraphics[width=.46\textwidth]{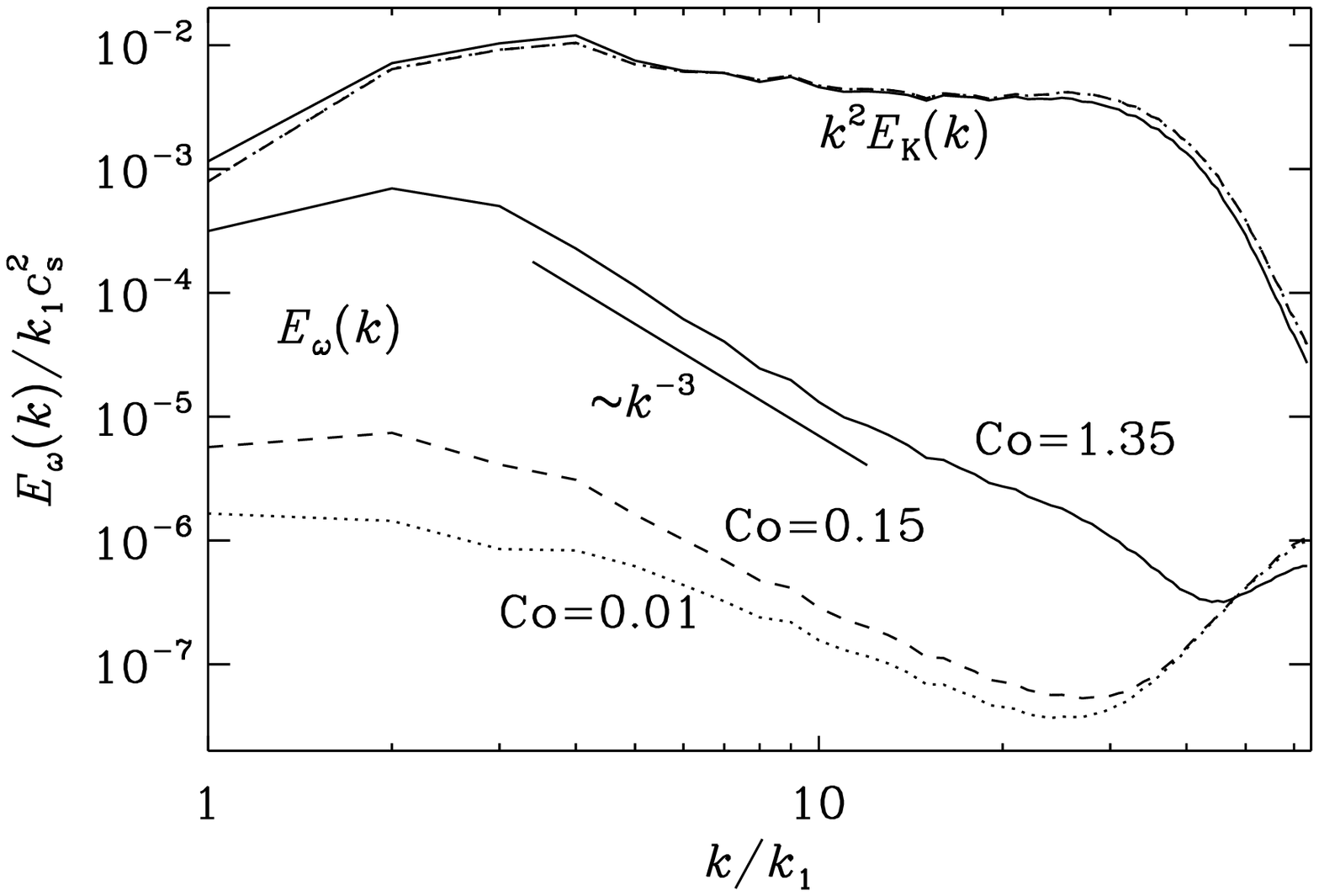}
\end{center}\caption[]{
Left: spectra of kinetic energy and enthalpy for two values of
the Coriolis number for $\Rey=25$ and $\St_{\rm force}=0.4$.
The two straight lines give the slopes $-2$ and $-3$, respectively.
Right: Enstrophy spectra, $E_\omega(k)$, compared with $k^2E_{\rm K}(k)$,
for $\Rey=25$, $\St_{\rm force}=0.4$, and three values of the Coriolis number.
}\label{pspec_comp}\end{figure}

We find that significant vorticity is only being produced
when the Coriolis number, $\Co=2\Omega/\urms\kf$ is about unity;
see \Fig{oo} for $\Co=0.15$ and 1.35.
For both cases we show in the left-hand panel of
spectra of kinetic energy and enthalpy,
$E_{\rm K}(k)$ and $E_{\rm h}(k)$, respectively.
There is no clear inertial range, but in all cases
the energy spectra show a clear viscous dissipation range.
There can easily be spurious vorticity generation,
possibly still due to marginally sufficient resolution.
The possibility of a spurious vorticity is indeed verified by
the right-hand panel of \Fig{pspec_comp}, where we compare
enstrophy spectra at different Coriolis numbers.
Note that for large values of $\Co$, the enstrophy spectrum decays
like $k^{-3}$.
However, for smaller values of $\Co$ the level of enstrophy at the
mesh scale remains approximately unchanged and is thus responsible
for the spurious vorticity found above for small values of $\Co$
and not too small values of $\Rey$.
For larger values of $\Co$, the production of vorticity
is an obvious effect of rotation in an otherwise potential velocity
field, and it is most pronounced at large length scales, as can also
be seen in the right-hand panel of \Fig{pspec_comp}.

In the presence of shear, we find, in analogy with the case of rotation,
production of vorticity proportional to the magnitude of the shear.
However our results indicate that under the typical physical conditions
in the interstellar medium in our Galaxy, neither rotation nor shear would
be strong enough to produce significant amounts of vorticity \citep{DSB10}.

\begin{figure}[t!]
\begin{center}
\includegraphics[width=\linewidth]{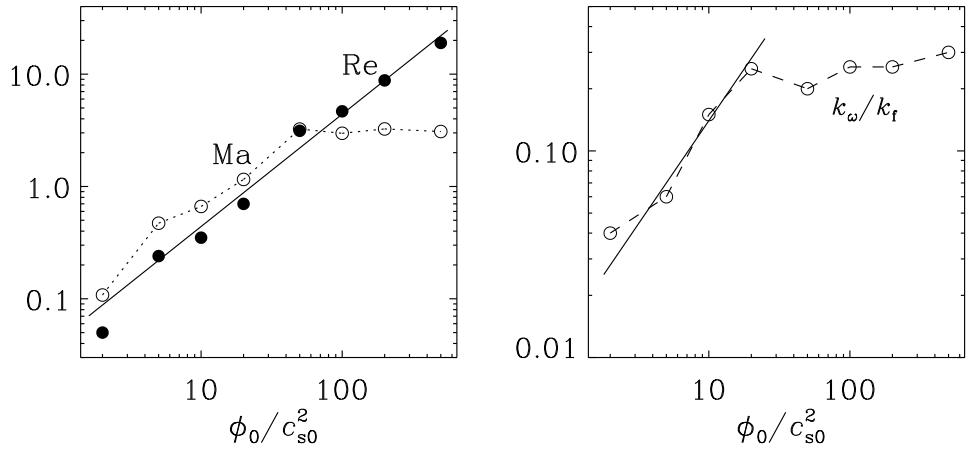}\qquad
\caption{
Dependence of $\Ma$ and $\Rey$, as well as the normalized vorticity,
$\kom/\kf$, on $\phi_0$ for $\nu/\cs R=1$.
}\label{pbaro_nu_3d}
\end{center}
\end{figure}

Finally, we relax the isothermal condition to let the system evolve
under the action of the baroclinic term.
In this situation we have non-parallel gradients of pressure and density.
The baroclinic term is proportional to the cross product of the two
gradients, resulting from the curl of the term $\rho^{-1}\nab p$.
In \Fig{pbaro_nu_3d} we show the dependence
of various quantities on the forcing amplitude $\phi_0$ normalized by
the reference sound speed $\csz$.
The Mach number saturates at about $\Ma=3$, and the rms value of
the entropy gradient increases up until this point.
The amount of vorticity production in terms of $\kom/\kf$
is about 0.3 for $\phi_0/\csz^2\ga20$.
For smaller values, on the other hand, there is an approximately
linear increase with $\kom/\kf\approx0.014\phi_0/\csz^2$.

Given that in our Galaxy the Mach number of the turbulence is about unity
\citep{Beck96} it is clear that the baroclinic term is much more efficient in
driving the production of vorticity.
The fact that the biggest amount of vorticity is observed when shock
fronts encounter each other suggests that supersonic conditions need
to be investigated more deeply.

Regarding dynamo action, as pointed out by \cite{BD09}, the presence of
vorticity does not seem to affect the diffusion of magnetic fields
differently than a complete irrotational turbulence.
Nevertheless, vorticity plays an important role in dynamo processes, so it is
important to address the problem of the generation of vorticity
and the possible role of other effects. 
In future work we will address the connection between vorticity generation
and the dynamo effect for magnetic fields.

\end{document}